# Efficient FPGA Floorplanning for Partial Reconfiguration-Based Applications

Norbert Deak
*Cluj-Napoca branch*

*National Instruments*
Cluj-Napoca, Romania
Norbert.Deak@ni.com

Octavian Creț
*Computer Science Department*

*Technical University of Cluj-Napoca*
Cluj-Napoca, Romania
Octavian.Cret@cs.utcluj.ro

Horia Hedeșiu
*Department of Electrical Machines,
Marketing & Management*

*Technical University of Cluj-Napoca*
Cluj-Napoca, Romania
Horia.Hedesiu@mae.utcluj.ro

*Abstract*— **Partial Reconfiguration (PR) is a technique that allows reconfiguring the FPGA chip at runtime. However, current design support tools require manual floorplanning of the partial modules. Several approaches have been proposed in this field, but only a few of them consider all aspects of PR, like the shape and the aspect ratio of the reconfigurable region. Most of them are defined for old FPGA architectures and have a high computational time. This paper introduces an efficient automatic floorplanning algorithm, which takes into account the heterogeneous architectures of modern FPGA families, as well as PR constraints, introducing the aspect ratio constraint to optimize routing. The algorithm generates possible placements of the partial modules, then applies a recursive pseudo-bipartitioning heuristic search to find the best floorplan. The experiments showed that the algorithm's performance is significantly better than the one of other algorithms in this field.**

## I. INTRODUCTION

Partial Reconfiguration (PR) makes it possible to modify certain parts of a hardware architecture implemented in an FPGA chip while the remaining parts of that architecture are still running ([1]). The advantage consists in the reduction of the design's size, thus also reducing the cost and the power consumption. PR offers flexibility in the choices of algorithms or protocols for an application, along with security improvements and also the possibility of introducing FPGA fault tolerance. But it also poses a few design challenges, one of the most important being the manual floorplanning (FP), where the user must provide a placement in the FPGA chip for each reconfigurable partition. This can be difficult to achieve because the user should have deep knowledge of the structure of the FPGA. Furthermore, each placement should meet the modules' requirements in terms of resources, taking into account the different types of resources and meeting special PR design constraints. For a larger number of reconfigurable regions, this problem may be very difficult to solve manually.

Several solutions were proposed for solving the FP problem, however only few of them consider PR and FPGAs with a heterogeneous architecture. In this paper, a new method that satisfies the PR constraints is proposed, which is aware of the heterogeneous distribution of resources and takes into account the inter-modules connections. This method uses the recursive pseudo-bipartitioning heuristics from [2] in order to reduce the problem size and also to find a convenient location for each module on the FPGA device. Instead of generating all the possible placements for each module as in [2], the Columnar Kernel Tessellation method [3] is used for rapidly generating the modules placement and for reducing the amount of wasted resources. Also, in the bipartitioning heuristics, the total wirelength is considered only indirectly for solving a Nonlinear Integer Programming (NLIP) problem [4]. A new constraint has also been introduced: the *Aspect Ratio* (AR) of the modules, which must be around 0.5 to avoid routing problems, as explained in [5], [6].

The FP problem works on a number of reconfigurable regions, each of them having its resource requirements. The target FPGA device is also chosen, as well as an objective function that needs to be minimized. The FPGA can be seen as a matrix, where the rows are the clock regions [7], [8]. Each cell contains a specific type of resource (mainly CLB, BRAM and DSP blocks). This is due to the fact that when PR is performed, the minimal reconfigurable unit is a column of 1 CLB wide and 1 clock region high. This minimal reconfigurable unit can be called a *tile*.

A solution to the FP problem is defined as a rectangular area for each region. The solution is valid if it meets these four constraints: (1) For each region, the chosen area must contain the resource requirements; (2) Overlapping between two regions is forbidden; (3) All regions should fit in the target FPGA chip; (4) The objective function of the problem is minimized (the resource wastage and the total wirelength).

There are also two additional, PR specific constraints that should be met. First, each region has to be placed into a rectangular area that includes complete tiles, and, second, the AR of a region has to be between 0.4 and 0.6 to ease the routing process. Authors of [5] claim that the best results are obtained for an AR of ~0.5, and the author of [6] claims that AR has to be less than 1 to ensure that the design is routable – but none of them provides a formal proof.

In FP, the area of the reconfigurable region is an important factor. PR allows changing the functionality of the region by uploading a partial bit file at runtime. This new uploaded configuration may require other types of resources or a different amount of them, for instance more CLBs or less DSPs, etc. This implies that at FP step, when selecting an area for the reconfigurable region, every partial module that may be implemented in that region has to be evaluated as regard to the amount of logic resources it needs, i.e. each module must have enough resources available. As a matter of consequence, the required amount of resources for the reconfigurable region is the minimal one that satisfies the resource requirements of each partial module that is implemented in the corresponding region.

The *resource wastage* for a particular area of a region is defined as the difference between the amount of resources of the actual reconfigurable area and the required one. The FP algorithm aims at minimizing this factor.

*Total wirelength* is another important factor which determines the quality of the floorplanner. The reconfigurable regions may be interconnected by signals, so it is important to place these modules as close to each other as possible. The total wirelength is computed as the sum of the distances between the modules, where the distance between two modules is multiplied by the number of signals that connect them. The distance between two modules is given by the distance between their geometrical centers.

The remainder of the paper is organized as follows: Section II provides a detailed description of the existing FP methods, emphasizing on their strengths and limitations. Section III presents the proposed bipartitioner, followed by Section IV, which describes the proposed algorithms in detail. The experimental results are presented in Section V, while Section VI concludes the paper.

## II. RELATED WORK

In the literature, several approaches to FPGA FP have been published. Most of them focus only on static placements, as they don't consider the partial modules. In [9] slicing trees are used to represent an FPGA floorplan and simulated annealing, one of the most common algorithms in FP [7], to find the optimal result. It takes into consideration that the FPGA contains CLBs, BRAMs and DSPs, but it relies on the fact that these resources are homogeneously distributed, as in older devices, while in newer FPGAs they are distributed heterogeneously.

HeteroFloorplan [10] is another FP for static modules which uses a slicing tree representation and relies on a three phase algorithm. The first step is the recursive balanced bipartitioning, followed by the slicing topologies generation. The final step implies running an algorithm based on a greedy heuristic and a min-cost max-flow formulation for allocating the resources for the modules. An improvement of the algorithm is presented in [11], however, both of them are based on the same FPGA architecture as the model used in [9].

In [3] a new method is presented, which focuses on partial reconfigurable modules and takes also into account the modern FPGAs' architecture. The algorithm considers the FPGA device as a collection of tiles containing a specific type and amount of resources (CLB tiles, BRAM tiles and DSP tiles). Each reconfigurable region is assigned a priority based on the type of resource needed, the requirements being expressed in tiles. Then, all kernels are created by merging adjacent tiles, and then they are extended horizontally until meeting the reconfigurable region requirements. This procedure is repeated several times using different initial kernels in order to find the best floorplan. At the end, a post processing step is performed to reduce the total wirelength by moving or swapping the allocated regions.

Another approach is proposed in [12], based on Mixed-Integer Linear Programming (MILP). It also considers the heterogeneously distributed resources existing in modern FPGAs along with the partial modules. An analytical model is proposed for the reconfigurable regions, using their resource requirements, their interconnections and special PR constraints.

A similar FP is presented in [2], which is aware of the heterogeneous distribution of resources and of the PR, but also takes into account all the components, their connectivity and resource requirements. First, the floorplanner finds all possible rectangular placements of each module, and then executes a recursive pseudo-bipartitioning using NLIP. Then the placement candidates are selected using the Pareto-ranking technique, and finally a trial-and-error algorithm is executed in order to obtain the final floorplan. This method is compared with the floorplanner from [12], showing that it performs better. Though, finding all the placements for the modules is intensively time consuming.

## III. PROPOSED APPROACH

Recursive bipartitioning is often used to solve the problem of FP [10]. The basic idea of a cut-size driven recursive bipartitioning is shown in Fig. 1. The circuit is cut in two partitions such that the cut line crosses as few wires as possible, also ensuring that the modules are balanced in the two new partitions. These cut lines can be vertical or horizontal. In Fig. 1, the connections between modules are shown as solid lines, the dashed lines being the cuts. In Fig. 1.a), a vertical cut is made intersecting a connection and separating the region in two halves. Likewise, in Fig. 1.b), the next two cuts are shown, which are also vertical, made recursively on the left, respectively on the right partition from the previous step (a).

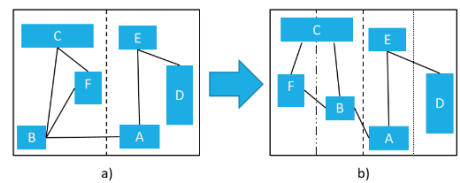

Fig. 1. The recursive bipartitioning example – [2].

A simple recursive bipartitioner, such as [10] is not possible for modern FPGAs because the resources are heterogeneously distributed. Therefore, in [2] a new bipartitioner was proposed, which can minimize the signal nets crossing the cut line and supports multi-resource bipartitioning. This means that the available resource types and quantities can differ in the two partitions, the resources occupied by the possible placements of one module in both partitions can be different, and the resources occupied by the modules in the partitions can be individually balanced.

The proposed method consists of using an NLIP-based bipartitioner mechanism (similar to [2]) to find the location of each module on the FPGA chip; however, we added new constraints to it in order to reduce the total wirelength. The partitioning problem is modeled here as an NLIP optimization problem. A binary variable, named $m_i$, shows the belonging of module $i$ to a partition module. The objective function to be minimized is the total number of nets that cross the cut line. We used the Manhattan metric to compute the distance between the regions. Equation (1) shows how to compute the number of nets that cross the cut line, where $n_{ij}$ is the number of signals linking the two modules $i$ and $j$:

$$nets_{ij} = n_{ij} \cdot (m_i + m_j - 2 \cdot m_i \cdot m_j) \qquad (1)$$

However, an additional constraint should be added. The wirelength is computed from the center of the modules, so its shape and size also matter when computing the distance between two modules, as illustrated in Fig. 2. This constraint can be implemented by first computing the average width and height of all the placements of the modules $i$ and $j$ in a partition, in both cases, starting with the placements in partition 0 and continuing with the placements in partition 1. Equation (2) shows the updated formula for computing the cut line, for a vertical cut, where the average width should be considered. $w_{i0}$ is the average width of all the placements of module $i$ in partition 0, $w_{i1}$ in partition 1, and the same for $w_{j0}$ and $w_{j1}$ (the formula for a horizontal cut is the same, except it considers the average height).

$$nets_{ij} = n_{ij} \cdot (m_i \cdot (w_{i1} + w_{j0}) + m_j \cdot (w_{i0} + w_{j1}) - m_i \cdot m_j \cdot (w_{i0} + w_{j0} + w_{i1} + w_{j1})) \qquad (2)$$

The objective function (the sum of the signal nets crossing the two partitions), is expressed in the equation (3):

$$\sum_i \sum_{j \neq i} nets_{ij} = \sum_i \sum_{j \neq i} n_{ij} \cdot (m_i \cdot W_1 + m_j \cdot W_2 - m_i \cdot m_j \cdot (W_1 + W_2)) \qquad (3)$$

Equation (3) contains the wirelength between the modules in the parent partition, where $W_1 = w_{i1} + w_{j0}$ and $W_2 = w_{i0} + w_{j1}$. But some of the modules have connections with other modules which are not in the parent partition; that connections could also cross the cutline, so they should be taken into consideration. Fig. 3 illustrates the differences for a corner case, when this situation is taken into consideration, and when it is not, respectively.

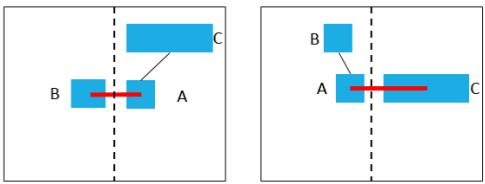

Fig. 2. The difference between wirelengths in the two cases: on the left, the distance between A and B is shorter than the distance on the right between A and C, because it is measured from the center of the modules.

Therefore, two additional module lists are constructed, one containing modules from the left side of the parent partition, and the other one from the right side. Then, using equation (2), we compute the connections between the modules in the parent partition, and then we compute the connections between those modules and the outside world. Finally, we sum them using equation (3). Instead of these two variables, a single one is used because we already know the location of the outside module (0 for the left, 1 for the right).

The bipartitioner's goal is to minimize the left-hand side of equation (3) taking into account the constraints. The total amount of resources assigned to the modules in a partition should not exceed the amount of available resources in that partition.

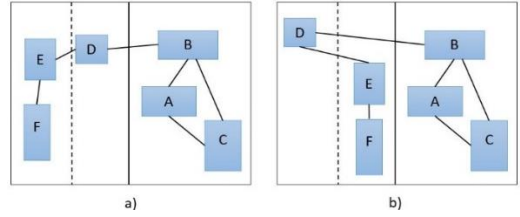

Fig. 3. The connection to modules that are not in parent partition. In (a), the current parent partition is the left part of the region, separated with a solid line, containing modules D, E, F. By considering the connection between D and B, it is more convenient to put D in the right partition, rather than in the left one (separated with dashed line). In (b), this connection is not considered, and it may end up putting D in the left one, increasing the number of connection that will be cut, thus increasing the wirelength.

For the CLB resources, these constraints are shown in equations (4) and (5). They have to be ensured for both partitions. $Total0_{CLB}$ shows the total number of CLBs occupied by the modules assigned to partition 0, and $Total1_{CLB}$ for partition 1. The available number of CLBs in the partitions is represented by $MAX0_{CLB}$, and $MAX1_{CLB}$ respectively. The number of CLBs occupied by module $i$ is $CLB0_i$ in partition 0, and $CLB1_i$ in partition 1.

$$Total0_{CLB} = \sum_i ((1 - m_i) \cdot CLB0_i) \leq MAX0_{CLB} \qquad (4)$$

$$Total1_{CLB} = \sum_i (m_i \cdot CLB1_i) \leq MAX1_{CLB} \qquad (5)$$

All constraints above have to be formulated not only for the CLBs, but also for the DSPs and BRAMs. To solve the minimization problem, we use an off-the shelf software application, Gurobi Optimization [13], [14].

## IV. IMPLEMENTATION

The first step is to create an empty root partition that will be used by the recursive bipartitioner. Next, all possible placements for each module are collected in a list, and then the recursive bipartitioner is executed twice, once horizontally and once vertically. When finished, for each placement of each module the normalized wastage and the distance from anchor point are computed. Then, the placements are sorted based on an objective function. Finally, the modules are sorted in the decreasing order of their size, and a feasible floorplan is found using trial-and-error.

In [2], the placement generation step is the most expensive one, as it computes all possible placements, disregarding the resource wastage, and it filters out large wastage regions in a future step. To avoid sorting and filtering a very long list, a smaller list of placements is generated, which contains only the placements which are efficient from the PR point of view: placements that contain only complete tiles, placements that are within the AR boundaries and placements that minimize the resource wastage.

This was implemented using the Kernel Tessellation Method [3], which generates all possible placements, and it does not consider the already chosen tiles, because the placements are not final, they are just a large list of possibilities. This step is presented in Algorithm 1. The algorithm works by prioritizing the resource requirements in the modules, since the FPGA has the least amount of DSP blocks, then BRAMs, and the majority of the resources are CLB tiles. Using this order, each module has primary, secondary, and possibly tertiary resource requirements.

---

**Algorithm 1.** Generate all possible placements

**Requires**: *modules* **and** *FPGA*
Filter the modules in four different lists: ($S_1$, $S_2$, $S_3$, $S_4$)
**For** *L* **in** the lists ($S_1$, $S_2$, $S_3$, $S_4$)
   Sort *L* based on the modules resource requirements
   **For each** module *M* in *L*
     Create all kernels in a new list *KL*
     Sort *KL* in increasing order of the kernels size
     **For each** kernel *K* in *KL*
        Expand *K* vertically until it has enough primary resources
        Expand *K* horizontally until it has enough secondary and tertiary resources
        Insert the expanded kernels in the placements list *PL* of module *M*
     **End for**
   **End for**
**End for**

---

First, four lists are generated and populated with modules, based on the following criteria: the first list $S_1$ contains the modules that include both DSP and BRAM resources, and possibly CLB tiles; the modules from the second list $S_2$ require DSP and also possibly CLB tiles, while modules from the third list $S_3$ include only BRAMs and possibly CLBs. Finally, the list $S_4$ comprises the modules containing only CLBs.

Then, for each list a number of steps are performed, starting by sorting the modules in ascending order based on their resource requirements. The lists are processed in this order: $S_1$ (DSP + BRAM + CLB), $S_2$ (DSP + CLB), $S_3$ (BRAM only), and finally $S_4$ (CLB only).

Resource requirements are ordered by their priorities. In the first list, $S_1$, the modules need both DSPs and BRAMs, and additionally also CLBs – this means that for the modules from $S_1$, the primary resource is DSP, the secondary one is BRAM, and finally, the tertiary resource in the priority list is CLB. In the next two lists, $S_2$ and $S_3$ (with DSP and CLB requirements, and BRAM and CLB requirements, respectively), the primary resource is DSP for $S_2$, and BRAM for $S_3$, but the secondary resource is CLB. There are no tertiary resources. Finally, in $S_4$, the only primary resource is CLB.

The kernels search starts from the bottom of the FPGA, from left to right. At the start of the row of the FPGA matrix (the clock region), the first primary resource tile is located. In case the kernel contains both primary and secondary resources, the nearest secondary resource is located and a new kernel is created. Otherwise, the kernel will contain only the primary resource tile. Then the next primary tile is located.

A kernel is valid only if it contains available tiles: no tile from the kernel is a reserved area (non-reconfigurable tiles). This is done at the level of the whole row until no more primary resource columns can be found. In case more primary resource tiles are needed, the algorithm expands these kernels on the vertical. But there can be placements that have multiple primary resource columns in a row.

To ensure that no possible placement is left out, these simple kernels are merged, creating larger kernels, which also include the other resources lying between the basic kernels. Fig. 4 illustrates this step in two cases: one where the primary resource is BRAM, and the other one, where the primary resource is DSP and the secondary one is BRAM (so, in this last case, a small kernel).

It is important that these kernels have as little resource wastage as possible, so if the modules already fit in the kernel, there is no need to merge the kernels, and the algorithm processes the next row. Otherwise, the kernels from the current row are merged in all possible ways. This is done by taking the first kernel in the list, which contains the kernels sorted out according to the FPGA device, and merging it with all the others, obtaining larger kernels. The algorithm stops at the kernel that has enough primary resources for the module. Then, this is done for all the next kernels in that row. Kernels are only valid if they do not contain reserved areas.

Next, all the other rows are processed in the same way. Finally, the algorithm returns the list of valid kernels, together with the merged kernels, from all rows. After all the kernels were generated, they are sorted based on their size, analyzed and processed, starting with the smallest ones. If a kernel doesn't have enough primary resource tiles, it will be expanded in a columnar way, i.e. vertically.

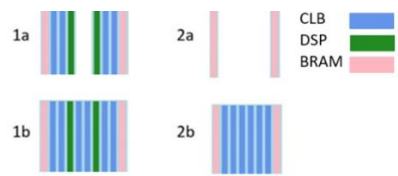

Fig. 4. An example of merging kernels. 1a shows two single, small kernels containing both DSP and BRAM, and they are merged in 1b, containing also the CLB tiles between them. 2a shows two small kernels, containing only BRAM. In 2b they are merged by also including them.

Since the kernel list contains all the kernels from the FPGA, the only way to avoid duplicates is to expand the kernel vertically until it gets the required primary resources. If it still needs more primary resources, then a larger kernel needs to be expanded, so the current one is discarded.

After it is ensured that the kernel has obtained the required primary resource tiles, the secondary requirements (if any) are analyzed. If it needs more resources, the kernel will be expanded horizontally, towards the secondary resource columns. The horizontal expansion is described in Algorithm 2. However, it does not grow only horizontally, but also upwards. This will solve some corner cases, where the module needs a very small amount of primary resources, but a large amount of secondary resources. The kernel will not grow vertically because it has already enough primary resources, but it may not get the needed secondary resources if it only grows horizontally.

With each new row added to the kernel after the vertical expansion, its height is computed, and based on it the algorithm computes how many secondary resource columns of equal height are needed (because of the AR constraints).

The kernel can be expanded horizontally in both directions. As shown in Fig. 5, every possible growth is taken into consideration as a new kernel. For instance, if 5 secondary columns are needed, then it first takes 5 secondary resource columns on the right, then 4 on the right, 1 on the left, and this continues until all 5 columns are taken from the left. By taking the next secondary resource column, all resources between the current kernel and the next secondary resource column are merged.

Also, if by merging a resource column the kernel would get a reserved area, it is discarded. Only those columns are added that would not merge other primary resource columns with the actual kernel, because that would be resulting in a larger kernel and duplicates must be avoided. This way it is ensured that the placement has enough secondary resources.

The horizontal expansion procedure is executed multiple times (see Algorithm 1). First, it is called when the module has the required number of primary resource tiles, but not enough secondary resources. Then, each new kernel that has enough secondary resources will be expanded horizontally until it gets enough tertiary resources. The same steps are repeated for the tertiary resources. Then, the generated kernels, which have enough resources, will be added to the modules placement list.

---

**Algorithm 2.** Horizontal kernel expansion

**Requires**: *kernel* **and** *module* **and** secondary resource
**Returns**: kernel list *KL*
**Repeat**
   <u>Get</u> the number *N* of secondary resource tiles needed by *module*
   **For each** combination of *l + r = N*
      <u>Create</u> new kernel *NK* by adding *l* secondary resource tiles to *kernel* from its left
      <u>Add</u> *r* secondary resource tiles to *NK* from the right of *kernel*
      <u>Insert</u> *NK* in *KL*
   **End for**
**Until** *kernel* cannot be expanded upwards

---

Before marking a kernel as a valid possible placement, it is also checked whether it respects the AR boundary constraint. All accepted kernels are saved in a list. An accepted kernel should meet the following constrains: (1) it does not include a reserved area; (2) its available resources contain at least the required resource for a module; (3) it is rectangular; and (4) it meets the AR constraints (the boundaries set for the AR are between 0.2 and 0.7, where the AR is defined as *width* over *height*. These boundaries were chosen by us based on empirical observations).

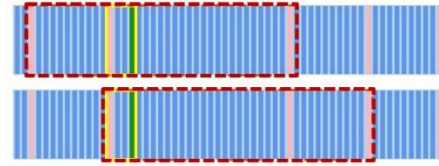

Fig. 5. Horizontal expansion of a kernel. The kernel is bordered with solid line, and the expanded kernel with dashed line. On the top, it takes one column to the left and one to the right, while on the bottom, two columns from the right, no column from left.

Next the recursive bipartitioner is executed on the root partition (the one that includes all the modules and their possible placements). This algorithm is executed twice, first performing a recursive vertical cut, and then a recursive horizontal cut. The goal of the bipartitioner is to find the location of each module, but instead of a final placement, it only estimates an *anchor point (geometrical center)*.

The algorithm creates two new partitions and aims at placing each module from the parent partition in one of the newly created ones. If the geometrical center of a placement is closer to the center of one of the child partition (at least 75% of a module fits into the partition), the module will be assigned to that partition. Otherwise, the module will remain in the parent partition. Fig. 6 illustrates the two cases.

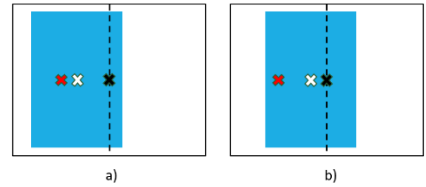

Fig. 6. The center of the parent partition, left and right partions are marked with a black, a red and a white corss, respectively. In (a), 90% of the module fits in the left partition, so its center is closer to the center of the partition. In (b), only 75% of the module fits, so its center is closer to the center of the parent partition.

An example of this process is illustrated in Fig. 7. It can be seen module A doesn't fit in any of the partitions, so it will be left out from further cuts, but still kept in the parent partition. Its resources must be also deducted from the partitions' available resources. The other modules, B, C, D, E and F fit into the partitions, so they will be assigned to one of the two partitions based on the cost function in the NLP model. Here, the AR of a placement cannot be changed, since for each placement of a module was already fixed in Algorithm 1.

To decide which module will belong to which partition, an NLIP model is built, based on the equations from the previous section, and then solved by the Gurobi Optimizer tool [13]. After every module has been assigned to a partition, their anchor points are updated with the geometrical central of the partition they belong to.

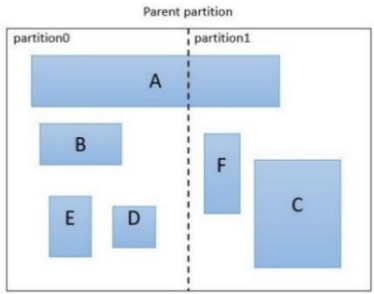

Fig. 7. The parent partition cut in half with a vertical cut (dahsed line). The modules will be assigned to one of the partition based on the constraints and the objective function.

In the end, if a valid solution is found, the recursive bipartitioner is executed for both partitions, but only if a partition contains at least one module. The recursion stops when all the partitions lists are empty.

After two recursive bipartitioning calls (horizontal and vertical), the normalized wastage and distance from the anchor point is computed for all the placements of each module. First, for each module the maximum distance from the anchor point and the maximum wastage are computed, and then for each placement, the distance from the anchor point is normalized. The same way, for each placement, the resource wastage is also normalized. In our method, no additional filtering is needed for the placements, as in [2], because the placements were generated to contain the minimal resource wastage. The main objective of the recursive bipartitioning algorithm is to minimize the cut-size so that the connected modules are closer to each other. By choosing the placements closer to the anchor point specified by the bipartitioner, the wirelength is minimized.

In this case, the weights of the wirelength and the resource wastage, specified at the beginning of the algorithm, can be applied directly to the distance from the anchor point and the resource wastage. This way, an objective value is obtained for each module, as shown in equation (6), and the placements for each module can be sorted based on this objective value. The *wastage* is the resource wastage of the placement, i.e. the difference of the actual resource in the placement and the required resource, while $\delta_A$ is the distance between the placement's center point and the module's anchor point (this indirectly represents the total wirelength).

$$OBJ_{placement} = \alpha \cdot wastage + \beta \cdot \delta_A \qquad (6)$$

The next step is to sort the modules in decreasing order of resource requirements. By doing this, in the next step, the modules with the highest requirements get placed, occupying most of the resources from the FPGA device, and this way, the module from the end of the list (the ones with the smallest resources) can easier and faster find a place between the large modules.

Finally, a trial-and-error algorithm is run to obtain the final placement of each module on the FPGA device. Having ordered the modules in the previous step, the algorithm doesn't need to backtrack all possible combinations of small modules, because they have a large number of possibilities. Therefore, it is easier to find a location for them next to the already placed large modules. The algorithm stops at the first feasible floorplan, which is found relatively fast due to the prior optimizations.

V. TESTING AND RESULTS

All experiments were performed on a 2.5 GHz Intel Core 2 Duo processor with 4 GB RAM under Windows 10, using Xilinx' Vivado software.

There are several problems when trying to compare this algorithm with previously proposed ones. Most of the papers published in the literature present their comparisons with other methods in form of percentages (i.e. "our algorithm is $x$% faster than another algorithm"). Another issue is the difference between FPGA architectures. Devices from the Virtex 5 family are rather small, having only one or two DSP48 columns. If executed for this family, our algorithm would exclude many possible placements; that is why it should be specially adapted for this type of devices. But devices from Virtex 4, Virtex 6 and Kintex 7 families have the BRAMs and DSPs relatively uniformly distributed all over the chip. So our algorithm shows its whole potential on these architectures.

We conducted the experiments on all the latest Xilinx families, Kintex7 and Virtex7 [15]. As the algorithms previously reported were tested on Virtex5, which was not our primary target as it is already an "old" family of FPGAs, we preferred making a complexity comparison.

*A. Performance estimation*

The running time of the FP algorithm depends on the size of the FPGA and the number of modules. The following parameters need to be finely tuned in our solution: $N$ – the number of modules; $R$ – the number of rows (clock regions) on the FPGA chip; $C$ – the number of columns; $D$ – the number of DSP columns; $B$ – the number of BRAM columns.

The first step is to create a root partition that represents the whole FPGA device, which parses all the tiles once; that is $O(R \cdot C)$. Next, the placements for the modules are generated: first, the modules are filtered in four different lists, which takes $O(N)$. The next steps are computed for each of the four lists separately, as they constitute different cases and cannot be generalized from the performance point of view. The estimated size of the first list $S_1$ (that contains modules with DSP, BRAM and possibly CLB requirements) is $N / 4$, so the first step's complexity (sorting) is $O(N \cdot log(N))$.

The next step consists of generating all base kernels. The following steps require both DSPs and BRAMs. The entire algorithm is then repeated $R$ times. To find the kernels containing DSP and BRAM tiles, the complexity is $D \cdot (O(C/B) + O(C/B))$, since there are $D$ DSP tiles in a row, and the distance to the nearest BRAM column from each DSP tile is on average of $C / B$. The number of generated kernels is $D$, so adding $D$ kernels to a new list takes $O(D)$. The merging takes longer: when a module needs more DSP tiles than the ones available in a clock region, the generation of all merged kernels is a double nested *for* loop of size $D$, giving $O(D^2)$, creating $D^2$ merged kernels that will be copied into the list. There are $R \cdot D^2$ kernels in the list, while the total runtime for generating the kernels is $R \cdot (D \cdot O(C/B) + O(D) + O(D^2))$, resulting in: $O(R \cdot (C + D^2))$.

The returned list is then sorted in $O(R \cdot D^2 \cdot \log(R \cdot D^2))$. For each kernel, in the worst case, a horizontal expansion must be made twice, once for BRAM and once for CLB, and its running time is $O(col^2)$, where $col$ is the number of columns needed by a resource. Both times the expansion stops at the next DSP tile, so for expanding to BRAM tiles, it requires at most $C / D$ columns. These can be expanded to add CLB tiles until the same number of columns: $C / D$. So, the complexity of the expansion is: $O\left(\left(\frac{C}{D}\right)^2\right)$.

The final complexity of generating all the placements for the first list is: $O\left(N \cdot \left(\log N + R \cdot (C^2 + D^2 \cdot log(R \cdot D^2))\right)\right)$. The average number of generated kernel for a module from the first list is: $R \cdot D^2 \cdot \frac{C}{D} \cdot \frac{C}{D} = R \cdot C^2$.

Similarly, one can compute the complexity for all the other lists. For the second list $S_2$ (modules needing only DSP and CLB resources), the complexity is the same. For the third list $S_3$ with only BRAMs and CLBs: $O\left(N \cdot \left(\log N + R \cdot (C^2 + B^2 \cdot log(R \cdot B^2))\right)\right)$. Finally, for the list $S_4$ containing only CLB based modules: $O\left(N \cdot \left(\log N + R \cdot C^2 \cdot log(R \cdot C^2)\right)\right)$. For all the three lists one can compute the same average kernel per module, $R \cdot C^2$.

The next step of the FP algorithm consists of the two calls of the recursive bipartitioning. The first step, creating two partitions from the parent partition is $O(R)$ and $O(C)$, depending on the cut type. For each module, its list of placements is parsed, giving $O(N \cdot R \cdot C^2)$. Finally, there are two recursive calls:

$$T(N, size) = 2 \cdot T\left(\frac{N}{2}, \frac{size}{2}\right) + O(N \cdot R \cdot C^2) + O(size) \quad (7)$$

where $size$ is $R$ or $C$, depending on the cut type.

### B. Complexity comparison with other methods

This section compares the complexity of the proposed solution with state-of-the-art works. In [2], the generation of all placements for the module is done by taking every possible rectangular placement on the FPGA. The complexity of this step is $O(N \cdot R^3 \cdot C^3)$. The number of placements generated for a module is $R^2 \cdot C^2$. Later, this list is processed multiple times. The algorithm can be run at most $C / 2$ times, until it finds a feasible solution. In this case, the complexity of the algorithm, except the placement generation, is multiplied by $C / 2$.

Table I reports the computational complexity of the proposed algorithm, compared to [2]. It can be noticed that the current solution has smaller running times for both steps of generating the placements and the recursive bipartitioning to find the anchor point (and finally find a feasible floorplan). Also, the number of placements for each module is smaller, and the placements contain the minimum amount of wastage.

TABLE I. COMPARISON OF THE ALGORITHMS' COMPLEXITY

| | PRFloor [2] | This algorithm |
|---|---|---|
| Placements generation complexity | $O(N \cdot R^3 \cdot C^3)$ | $O(N \cdot (\log N + R \cdot C^2 \cdot \log(R \cdot C^2)))$ |
| Placements per module | $R^2 \cdot C^2$ | $R \cdot C^2$ |
| Algorithm runs | $C/2$ | 1 |

### C. Performance comparison with other methods

Surprisingly, there are not many testing methods reported in the literature. Most of the existing benchmarking sets were designed for ASICs and only a few of them were transformed and adapted for FPGAs (like the MCNC and GSRC benchmarks used in [16], [17], and [10]).

The previously reported algorithms were tested either on pseudo-randomly generated circuits ([2], [12] and [16]) or on Software Defined Radios ([3] and [12]). As mentioned before, the comparisons with other methods presented by the authors of these papers are done only in relative terms (percentages) or the circuits themselves are relatively simple. We have chosen to use the same methods and benchmarks where available and we present the results below.

#### 1) Testing on Software Defined Radios

The algorithm was tested on a real case study taken from [3]. The design considered is a Software Defined Radio (SDR), consisting of a chain of modules, a matched filter, carrier recovery, demodulator, signal decoder and video decoder. Each module can have multiple modes that can be interchanged using PR. So, each module will have a reconfigurable partition, and each mode corresponds to a partial module. However, the FP is concerned only with the partitions. All modules are connected in sequential order with a 64 bit wide bus. The regions for the modules and their resource requirements are given in Table II. The target device is a Virtex 5 FX70T FPGA, containing CLB, BRAM and DSP tiles (each of them contains 36, 30 and 28 frames respectively).

TABLE II. RESOURCE UTILIZATION OF EACH MODULE IN THE SDR DESIGN

| Benchmark module | CLB | BRAM | DSP | No. of Frames |
|---|---|---|---|---|
| Matched Filter | 500 | 0 | 34 | 1040 |
| Carrier Recovery | 123 | 0 | 8 | 280 |
| Demodulator | 97 | 8 | 0 | 240 |
| Decoder | 234 | 2 | 0 | 462 |
| Video Decoder | 1100 | 6 | 34 | 2180 |
| **TOTAL** | **2054** | **16** | **76** | **4202** |

We used the same objectives as in [3], i.e. the minimum wirelength and the minimum wastage. The same circuit is tested in [12], so the comparison includes that result too. In both [3] and [12], a constraint specific to PR regions is not met, which is to keep the AR in a specific range, preferably [0.4-0.6], in order to avoid routing problems. We aimed at meeting this constraint, but during the experiments we noticed that the best interval is [0.2-0.7]. We performed the tests both with meeting the AR constraint and not. Table III presents the comparative results. We aimed at finding the optimal results by testing each possibility. The results for the area wastage were 306 when the AR constraint was not taken into consideration and 476 otherwise. In our FP algorithm, these optimal results are achieved in case the area wastage weight is set to the maximum. However, the best case is the balanced one, when neither the wastage nor the wirelength is too large.

From [3], there is no information about the running time, but in [12], the algorithm ran in this case in 29 seconds. For this design, our algorithm runs every time under a second.

TABLE III. RESULTS PROVIDED BY VARIOUS FP ALGORITHMS

| Algorithm | Benchmark module | Wastage | Wirelength | Runtime (s) |
|---|---|---|---|---|
| [3] | Min wastage | 466 | 8640 | - |
| [3] | Min wirelength | 516 | 7392 | - |
| [12] | Min wastage | 306 | ~8600 | 29 |
| This | Balanced | 504 | 11584 | 2 |
| This | Min wastage | 476 | 16704 | 1 |
| This | Min wirelength | 562 | 11328 | 1 |
| This | No AR, balanced | 506 | 10432 | 1 |
| This | No AR, min wastage | 306 | 8448 | 1 |
| This | No AR, min wirelength | 506 | 10432 | 1 |

*2) Testing on Pseudo-Randomly generated designs*

The performance of our algorithm was also tested on a set of pseudo-random circuits, on a Kintex 7 XC7K410T device using both the wirelength and the resource wastage, with the same weight, as objective function. The main idea for the pseudo-random circuits was described in [12], [18] and [16], but it had to be modified to show the true potential of this algorithm on latest families FPGA devices.

A series of pseudo-random circuits were generated with 5, 10, 15, 20, 25, 35, and 50 reconfigurable partitions, respectively. In each case, different types of circuits were generated, having the occupancy rate of 70% or 80%. To ensure that the circuits are heterogeneous, some modules require BRAM, and some require DSP blocks. Further, two modules are connected by a 64 bits wide signal with a probability of 1 / *number_of_modules*. Table IV reports the details of the experiment.

TABLE IV. EXECUTION TIME IN DIFFERENT SYSTEM CONFIGURATIONS.

| No. of PR Regions | CLB | BRAM | DSP | Runtime (s) |
|---|---|---|---|---|
| 5 | 70% | 5% | 3% | 51 |
| 10 | 70% | 11% | 6% | 79 |
| 15 | 70% | 13% | 8% | 83 |
| 20 | 70% | 16% | 9% | 101 |
| 25 | 70% | 27% | 19% | 61 |
| 35 | 70% | 37% | 29% | 52 |
| 50 | 80% | 30% | 30% | 37 |

*3) Testing on MCNC and GSRC benchmarks*

Since there is no standard benchmark available for FPGA FP, we have used some circuits from MCNC and GSRC suites for VLSI design and transformed them to cope with the FPGA FP scenario, similarly to [10], [16] and [17]. The main steps of this transformation are described in [16]. The results are presented in Table V.

TABLE V. RESULTS ON MCNC AND GSRC BENCHMARKS

| Circuit | No. of Reconfigurable Areas | Wirelength | Execution time (s) |
|---|---|---|---|
| apte | 9 | 11009 | 4 |
| xerox | 10 | 60676 | 10 |
| hp | 11 | 9670 | 4 |
| ami33 | 33 | 125262 | 110 |
| n10 | 10 | 2934 | 15 |
| n30 | 30 | 7123 | 10 |

## VI. CONCLUSIONS AND FURTHER WORK

This paper presents a new approach for FP in FPGA devices with PR capabilities. Previous works reported in the literature were based on older FPGA architectures, which have significantly changed over time, and the algorithms had large execution times even for the smallest design.

The main contribution of this paper is the introduction of a high performance algorithm (especially from the execution time point of view), which can be used on all FPGA architectures, including the newest ones. It is based on an architecture aware algorithm to generate all the possible placements with a minimum wastage of FPGA resources. The best placement is chosen using a bipartitioner, whose main goal is to minimize the total wirelength.

As there are no standardized benchmark designs for testing this category of algorithms for FPGAs, the theoretical performance of the proposed algorithm was thoroughly analyzed and compared with previous algorithms from the literature. Generating all placements is the most time consuming step of this type of algorithms. In our algorithm, the complexity of this step is lower than the one exhibited by other algorithms; it has a very significant contribution (more than 90%) to the reduction of the execution time. Also, the number of placements per module is smaller.

The tests were executed such that they provide as many fair comparisons as possible. Our algorithm shows its whole potential on the latest FPGA architectures. Unlike previous approaches, this algorithm considers the aspect ratio constraint of partial modules to avoid routing problems. The experiments with different system setups show that, in most of the cases, a result can be generated much quicker than previously reported similar algorithms.

Future work will focus on further reducing the wirelength and integrating the algorithm with existing commercial tools, making a partial reconfiguration design flow fully automatic.